\documentclass[a4paper]{article}
\usepackage{issp2024_abstract,amssymb,amsmath,graphicx}
\usepackage[utf8x]{inputenc}
\usepackage[T1]{fontenc}
\usepackage{wasysym}
\usepackage[english]{babel} 
\usepackage[autostyle]{csquotes}
\usepackage{caption}

\title{Discovering phoneme-specific critical articulators through a data-driven approach}

\makeatletter
\def\name#1{\gdef\@name{#1\\}}
\makeatother
\name{{Jesuraj Bandekar, Sathvik Udupa, Prasanta Kumar Ghosh}}

\address{\small \em Electrical Engineering Department, Indian Institute of Science, Bangalore\\
{\small \tt jesurajbandekar.661@gmail.com, sathvikudupa66@gmailcom, prasantg@iisc.ac.in}}

\begin{document}
\maketitle


\vspace{-1.4cm}
\section{Introduction}
Speech articulators such as tongue, lips, jaw and velum play an important role in speech production. The varying positions of these articulators result in sounds referred to as phonemes. It is known that, for different phonemes, certain articulators consistently achieve their target positions, and these are known as critical articulators \cite{JACKSON2009695}. The non-critical articulators' positions tend not to directly affect the characteristics of the phoneme produced.    
Let us consider the use of time-varying articulatory trajectories obtained from Electromagnetic articulography (EMA); the presence of non-critical articulators may be a source of noise for various modelling tasks such as Acoustic to Articulatory Inversion (AAI), Articulatory to Acoustic Forward (AAF) mapping, Phoneme to Articulatory (PTA) mapping etc. Thus, we believe that measuring the degree of criticality of articulators for different phonemes is beneficial. There has been previous work by \cite{anusuya2020data} which measures the criticality of articulators using the data distribution of articulatory movements. In this work, we are interested in exploring if such observations can be discovered in an unsupervised manner while training a neural network model for phoneme classification task. 

\vspace{-0.4cm}
\section{Methods}
We propose to learn phoneme-specific critical articulators in an unsupervised fashion through a data-driven end-to-end machine-learning approach. To achieve this, we use two blocks of neural networks to perform three tasks - AAI; \cite{liu2015deep, illa2018low}, AWP (articulator weight prediction) and FPC (frame-level phoneme classifier). These three tasks are learned end-to-end with speech features as the input and frame-level phonemes as the final output, along with time-varying articulatory features and time-varying articulatory weights as intermediary outputs. We perform AAI using a neural network, mapping input 13-dim MFCC features to 12-dim articulatory trajectories, $x$ and $y$ coordinates of six articulators: $UL$, $LL$, $JAW$, $TT$, $TB$, and $TD$. This 12-dim data is the intermediary output, which is trained with Mean squared error (MSE) loss with ground truth articulatory movements. For AWP, we use a similar neural network and a different dense output layer to predict 12-dim features from input MFCCs. These features are further normalised through min-max normalisation across the 12 features. This acts as normalised weights for the articulators. Finally, we multiply the normalised weights and articulatory features to weigh the articulators; this acts as the input to FPC. The FPC is another neural network which predicts frame-level phoneme labels. This is optimised using frame-level cross-entropy loss against ground truth frame-level phonemes. We use transformer-based neural architecture for all models, following the architecture used by AAI in \cite{udupa21_interspeech}.
\\

We use the network described above to allow the model to estimate phoneme-specific critical articulators based on the weights learnt, and we find all three tasks are necessary for this purpose. We need AWP to predict features that can act as weights for articulators, where higher weights represent critical articulators. However, AWP doesn’t have an objective function; it is learned unsupervised. For it to learn higher weights for critical articulators, it needs a task which assists this learning. Hence, we use FPC so that the FPC loss can guide the AWP network to learn the required features. Now, the need for an AAI network arises - theoretically, ground truth articulatory movements could be multiplied with AWP weights for the FPC network. However, during training, we find that having an AAI network to predict articulatory movements improves the accuracy of predicted weights. We hypothesise that this is the case as it benefits phone prediction through the FPC - AAI backward pass, allowing more access to MFCC features.
We add additional optimisations to achieve better weights from the AWP network. Firstly, after AAI, we use a straight-through estimator (STE) [\cite{bengio2013estimating}] to replace predicted articulators with ground truth articulators while maintaining gradient flow from FPC to AAI, i.e., STE acts as an identity function for the backward pass. Additionally, we use a dropout of 0.5 on the 12-dim weight prediction features in AWP. We find that, without dropout, the AWP weights tend to get activated for particular articulators for all phonemes. 
\\

To summarise, the goal of our model formulation is to learn frame-level weightage across articulators (AWP), where high scores represent critical articulators. We then add a task, i.e., FPC, where this information could be learnt unsupervised. Finally, we add various optimisation methods to improve the weight predictions from AWP (AAI, dropout, STE). We implement our network in PyTorch and train the model on data of around 5 hours from 10 subjects comprising both acoustic and articulatory data \cite{udupa21_interspeech}. The implementation is available on our public GitHub repository\footnote{https://github.com/coding-phoenix-12/CriticalArticulatorUSL}.

\vspace{-0.5cm}
\section{Results and Discussion}
\begin{figure}
\vspace{-0.5cm}
    \centering
    \includegraphics[width=16cm]{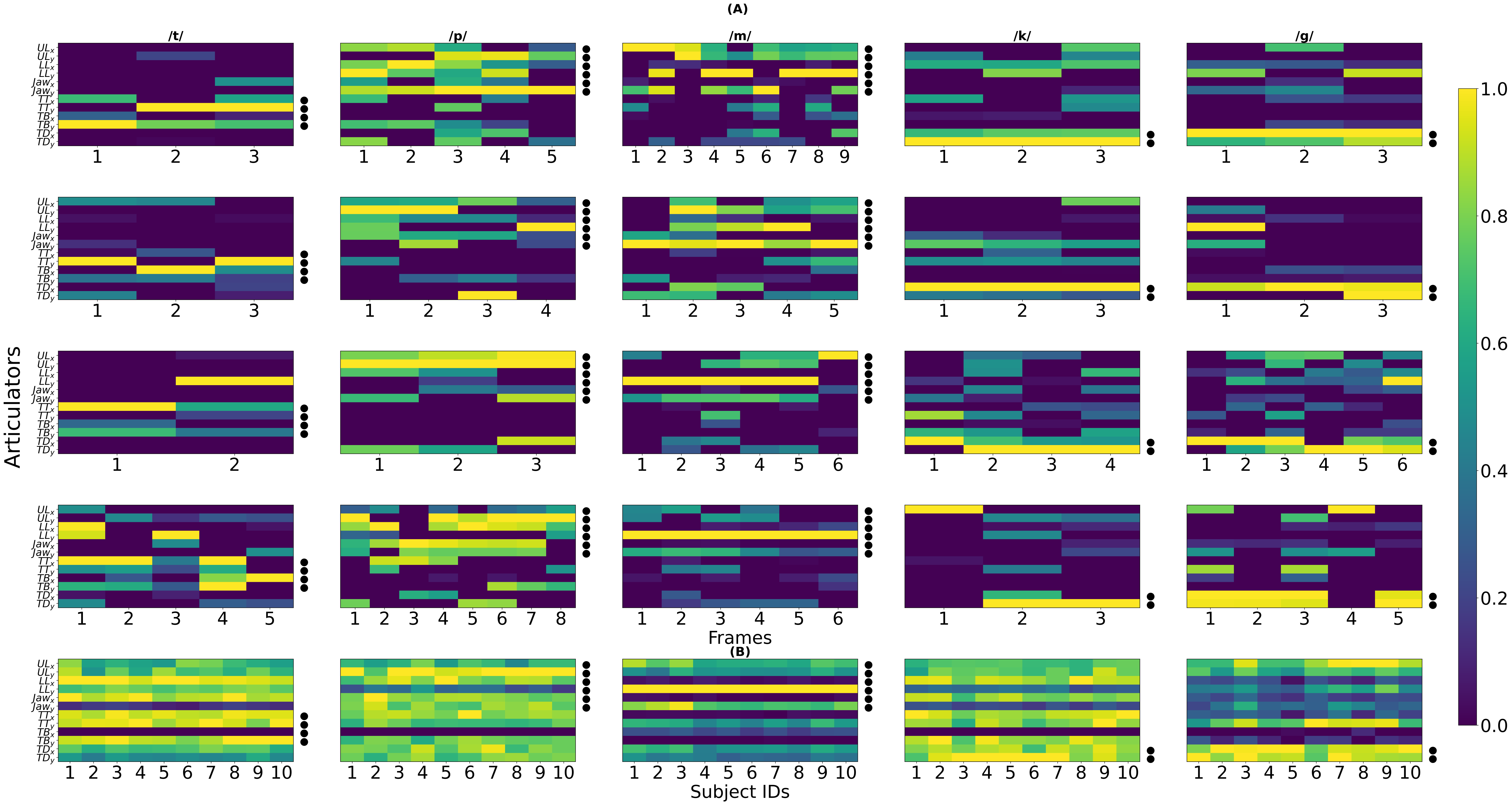}
    \vspace{-0.4cm}
    \caption{(A) The first four rows show the weights learnt for five phone segments  (the $x$-axis label indicates frame index) in different utterances. The articulators known to be critical are marked by \CIRCLE. (B) The last row represents the corresponding phoneme's average representation across all segments. It is shown for all 10 subjects  (the $x$-axis label indicates the frame index) used in this study.} 
    \label{fig:enter-label}
    \vspace{-0.4cm}
\end{figure}

Here, we analyse the effectiveness of the weights predicted for articulators. We visualise the normalised weights for five different phonemes (one phoneme in one column) for four segments as shown in Figure \ref{fig:enter-label}. Additionally, we also show the average representation of all phoneme segments in the last row. 
We can observe that the critical articulators \footnote{https://home.cc.umanitoba.ca/~krussll/phonetics/ipa/articulatory-ipa.html} are consistently activated for the relevant phonemes. For example, for \textbf{/t/}, we can observe that many tongue articulators are prominently activated. Additionally, it can be seen that the axis of the critical articulator can also learned in this process, such as activating only $x$ or $y$ axis. For example, for \textbf{/m/} $LL_y$ is activated, while $LL_x$ is not; this is in line with speech production as the movement along $y$ axis is necessary to form the constriction to produce \textbf{/m/}.  
This validates using our unsupervised approach to learn the weightage of critical articulators using an auxiliary task. Rather than grouping articulators as critical or non-critical, we get a measure of criticality from the weights. For example, for \textbf{/k/} we can observe other articulators, such as lips and tongue, are slightly activated; This could help in several tasks where even non-critical articulators can play a role.
Next, we use the speaker-level data from Figure 1(B) and compute the overall phoneme level average across articulators. These are the top three articulators activated for each phoneme - \textbf{/t/} : $LL_x$ $TT_y$ $TT_x$, \textbf{/p/} : $UL_y$ $TT_x$ $LL_x$, \textbf{/m/} : $LL_y$ $Jaw_y$ $UL_x$, \textbf{/k/} : $TD_y$ $TT_x$ $LL_x$ and \textbf{/g/} : $TD_x$ $TD_y$ $UL_x$. We can observe that for these five phonemes, there are at least 2 critical articulators activated in the top three, except for \textbf{/k/} where a single articulator is present. We believe that other phonemes are also activated due to co-articulation.

In the future, we will look at improving the weightage of articulators and identify the effect of co-articulation. Further, we plan to use the weights for different tasks involving articulatory trajectories.

\eightpt


\end{document}